\documentclass[11pt]{article}
\usepackage{graphicx}

\usepackage{amsmath}
\usepackage{epsfig}
\newcommand{\BABARPubYear}    {04}

\newcommand{\BABARConfNumber} {006}
\newcommand{\SLACPubNumber} {10642}

\RequirePackage{xspace}

\usepackage{relsize}
\def\babar{\mbox{\slshape B\kern-0.1em{\smaller A}\kern-0.1em
    B\kern-0.1em{\smaller A\kern-0.2em R}}}

\def\epem       {\ensuremath{e^+e^-}\xspace}

\def\piz   {\ensuremath{\pi^0}\xspace}

\def\pip   {\ensuremath{\pi^+}\xspace}
\def\pim   {\ensuremath{\pi^-}\xspace}
\def\pipi  {\ensuremath{\pi^+\pi^-}\xspace}

\def\Kbar  {\kern 0.2em\overline{\kern -0.2em K}{}\xspace}

\def\Kz    {\ensuremath{K^0}\xspace}
\def\Kzb   {\ensuremath{\Kbar^0}\xspace}
\def\KzKzb {\ensuremath{\Kz \kern -0.16em \Kzb}\xspace}
\def\Kp    {\ensuremath{K^+}\xspace}
\def\Km    {\ensuremath{K^-}\xspace}
\def\Kpm   {\ensuremath{K^\pm}\xspace}

\def\KpKm  {\ensuremath{\Kp \kern -0.16em \Km}\xspace}
\def\KS    {\ensuremath{K^0_{\scriptscriptstyle S}}\xspace} 
 
\def\Kstarz  {\ensuremath{K^{*0}}\xspace}
\def\Kstarzb {\ensuremath{\Kbar^{*0}}\xspace}

\def\Kstarp  {\ensuremath{K^{*+}}\xspace}

\def\Dbar    {\kern 0.2em\overline{\kern -0.2em D}{}\xspace}

\def\Dz      {\ensuremath{D^0}\xspace}
\def\Dzb     {\ensuremath{\Dbar^0}\xspace}
\def\DzDzb   {\ensuremath{\Dz {\kern -0.16em \Dzb}}\xspace}
\def\Dp      {\ensuremath{D^+}\xspace}
\def\Dm      {\ensuremath{D^-}\xspace}

\def\DpDm    {\ensuremath{\Dp {\kern -0.16em \Dm}}\xspace}

\def\Dstarz  {\ensuremath{D^{*0}}\xspace}
\def\Dstarzb {\ensuremath{\Dbar^{*0}}\xspace}

\def\Dstarm  {\ensuremath{D^{*-}}\xspace}

\def\B       {\ensuremath{B}\xspace}
\def\Bbar    {\kern 0.18em\overline{\kern -0.18em B}{}\xspace}

\def\Bz      {\ensuremath{B^0}\xspace}
\def\Bzb     {\ensuremath{\Bbar^0}\xspace}
\def\BzBzb   {\ensuremath{\Bz {\kern -0.16em \Bzb}}\xspace}
\def\Bu      {\ensuremath{B^+}\xspace}
\def\Bub     {\ensuremath{B^-}\xspace}
\def\Bp      {\ensuremath{\Bu}\xspace}

\def\BpBm    {\ensuremath{\Bu {\kern -0.16em \Bub}}\xspace}

\def\BorBbar    {\kern 0.18em\optbar{\kern -0.18em B}{}\xspace}
\def\DorDbar    {\kern 0.18em\optbar{\kern -0.18em D}{}\xspace}
\def\KorKbar    {\kern 0.18em\optbar{\kern -0.18em K}{}\xspace}

\mathchardef\Upsilon="7107
\def\Y#1S{\ensuremath{\Upsilon{(#1S)}}\xspace}% no space before {...}!

\mathchardef\Deltares="7101
\mathchardef\Xi="7104
\mathchardef\Lambda="7103
\mathchardef\Sigma="7106
\mathchardef\Omega="710A

\def\Deltabar{\kern 0.25em\overline{\kern -0.25em \Deltares}{}\xspace}
\def\Lbar{\kern 0.2em\overline{\kern -0.2em\Lambda\kern 0.05em}\kern-0.05em{}\xspace}
\def\Sigbar{\kern 0.2em\overline{\kern -0.2em \Sigma}{}\xspace}
\def\Xibar{\kern 0.2em\overline{\kern -0.2em \Xi}{}\xspace}
\def\Obar{\kern 0.2em\overline{\kern -0.2em \Omega}{}\xspace}
\def\Nbar{\kern 0.2em\overline{\kern -0.2em N}{}\xspace}
\def\Xb{\kern 0.2em\overline{\kern -0.2em X}{}\xspace}

\def\BR         {{\ensuremath{\cal B}\xspace}}

\def\mes        {\mbox{$m_{\rm ES}$}\xspace}

\def\DeltaE     {\mbox{$\Delta E$}\xspace}

\newcommand{\tev}{\ensuremath{\mathrm{\,Te\kern -0.1em V}}\xspace}
\newcommand{\gev}{\ensuremath{\mathrm{\,Ge\kern -0.1em V}}\xspace}
\newcommand{\mev}{\ensuremath{\mathrm{\,Me\kern -0.1em V}}\xspace}
\newcommand{\kev}{\ensuremath{\mathrm{\,ke\kern -0.1em V}}\xspace}
\newcommand{\ev}{\ensuremath{\mathrm{\,e\kern -0.1em V}}\xspace}
\newcommand{\gevc}{\ensuremath{{\mathrm{\,Ge\kern -0.1em V\!/}c}}\xspace}
\newcommand{\mevc}{\ensuremath{{\mathrm{\,Me\kern -0.1em V\!/}c}}\xspace}
\newcommand{\gevcc}{\ensuremath{{\mathrm{\,Ge\kern -0.1em V\!/}c^2}}\xspace}
\newcommand{\mevcc}{\ensuremath{{\mathrm{\,Me\kern -0.1em V\!/}c^2}}\xspace}
\def\invfb   {\ensuremath{\mbox{\,fb}^{-1}}\xspace}

\def\mus  {\ensuremath{\rm \,\mus}\xspace}

\def\mus        {\ensuremath{\,\mu{\rm s}}\xspace}    %% microsecond
\def\ra                 {\ensuremath{\rightarrow}\xspace}

\def\pep2{PEP-II}

\def\gsim{{~\raise.15em\hbox{$>$}\kern-.85em
          \lower.35em\hbox{$\sim$}~}\xspace}
\def\lsim{{~\raise.15em\hbox{$<$}\kern-.85em
          \lower.35em\hbox{$\sim$}~}\xspace}

\def\CP                {\ensuremath{C\!P}\xspace}
\xspace

\newcommand{\jprlBase}       {Phys.\ Rev.\ Lett.\xspace}
\newcommand{\jprBase}        {Phys.\ Rev.\xspace}
\newcommand{\jplBase}        {Phys.\ Lett.\xspace}

\newcommand{\zpBase}         {Z.\ Phys.\xspace}

\newcommand{\plb}       [1]  {\jplBase\ B~{\bf #1}}

\newcommand{\jprl}      [1]  {\jprlBase\ {\bf #1}}
\newcommand{\jprd}      [1]  {\jprBase\ D~{\bf #1}}

\newcommand{\progtp}    [1]  {{Prog.\ Theor.\ Phys.\ {\bf #1}}}

\newcommand{\zp}        [1]  {\zpBase\ {\bf #1}}
\def\jetset74   {\mbox{\tt Jetset \hspace{-0.5em}7.\hspace{-0.2em}4}\xspace}
\def\stwobg   {\ensuremath{\sin(2\beta+\gamma)}\xspace}
\def\KKstarz  {\ensuremath{K^{(*)0}}\xspace}
\def\KKstarzb {\ensuremath{\Kbar^{(*)0}}\xspace}

\def\DDstarz {\ensuremath{D^{(*)0}}\xspace}
\def\DDstarzb{\ensuremath{\Dbar^{(*)0}}\xspace}

\providecommand{\KPi}{\mbox{\ensuremath{K^-\pi^+}}}
\providecommand{\KPiPiPi}{\mbox{\ensuremath{K^-\pi^+\pi^-\pi^+}}}
\providecommand{\KPiPiz}{\mbox{\ensuremath{K^-\pi^+\pi^0}}}

\providecommand{\BR}{\mbox{\ensuremath{\mathcal B}}}

\newcommand{\optbar}[1]{\shortstack{{\tiny (\rule[.4ex]{1em}{.1mm})}\\ [-.7ex] $#1$}}
\def\BzorBzbar    {\kern 0.18em\optbar{\kern -0.18em \Bz}{}\xspace}
\def\BorBbar    {\kern 0.18em\optbar{\kern -0.18em B}{}\xspace}
\def\DorDbar    {\kern 0.18em\optbar{\kern -0.18em D}{}\xspace}

\def\fisher {\ensuremath{\mathcal F}\xspace}

\def\BBcount {124}
\def\offlumi {12}
\def\dzkzBrVal  {6.2}
\def\dzkzBrStat {1.2}
\def\dzkzBrSyst {0.4}
\def\dstarzkzBrVal  {4.5}
\def\dstarzkzBrStat {1.9}
\def\dstarzkzBrSyst {0.5}
\def\dzbkstarzBrVal  {6.2}
\def\dzbkstarzBrStat {1.4}
\def\dzbkstarzBrSyst {0.6}
\def\dzkstarzBrVal  {1.1}
\def\dzkstarzBrStat {1.1}
\def\dzkstarzBrSyst {1.2}
\def\dzkstarzLim    {4.1}
\def\brscale{\ensuremath{\times10^{-5}}}
\long\def\inst#1{\par\nobreak\kern 4pt\nobreak
    {\it #1}\par\vskip 10pt plus 3pt minus 3pt}

\begin{document}
{\pagestyle{empty}

\begin{flushright}
\babar-CONF-\BABARPubYear/\BABARConfNumber \\
SLAC-PUB-\SLACPubNumber \\
%hep-ex/\LANLNumber \\
August 2004 \\
\end{flushright}

\par\vskip 5cm

% Title of the paper
\begin{center}
\Large \bf A study of $\Bzb\ra\DDstarz\KKstarzb$ decays
\end{center}
\bigskip

\begin{center}
\large The \babar\ Collaboration\\
\mbox{ }\\
\today
\end{center}
\bigskip \bigskip

% Abstract
\begin{center}
\large \bf Abstract
\end{center}
We present a study of the  decays $\Bzb\ra\DDstarz\KKstarzb$
using a sample of \BBcount\ million $\Y4S\ra B \bar B$ decays
collected with the \babar\ detector at the \pep2\ asymmetric-energy
\epem\ collider at SLAC.
We report evidence for the decay of \Bz\ and \Bzb\  mesons to the $\Dstarz\KS$
final  state with an average branching fraction 
$\BR(B \ra\Dstarz\Kzb)
=(\dstarzkzBrVal\pm\dstarzkzBrStat\pm\dstarzkzBrSyst)\brscale$.
Similarly, we measure 
$\BR(B \ra\Dz\Kzb)=(\dzkzBrVal\pm\dzkzBrStat\pm\dzkzBrSyst)\brscale$
for the $\Dz\KS$ final state.
We also measure
$\BR(\Bzb\ra\Dz\Kstarzb)
=(\dzbkstarzBrVal\pm\dzbkstarzBrStat\pm\dzbkstarzBrSyst)\brscale$
and set a 90\% C.L. upper limit $\BR(\Bzb\ra\Dzb\Kstarzb) < \dzkstarzLim\brscale$.
All results presented in this paper are preliminary.
\vfill
\begin{center}

Submitted to the 32$^{\rm nd}$ International Conference on High-Energy Physics, ICHEP 04,\\
16 August---22 August 2004, Beijing, China

\end{center}

\vspace{1.0cm}
\begin{center}
{\em Stanford Linear Accelerator Center, Stanford University, 
Stanford, CA 94309} \\ \vspace{0.1cm}\hrule\vspace{0.1cm}
Work supported in part by Department of Energy contract DE-AC03-76SF00515.
\end{center}

\newpage
} % end of pagestyle{empty}

\begin{figure}[htb]
\begin{center}
\epsfig{file=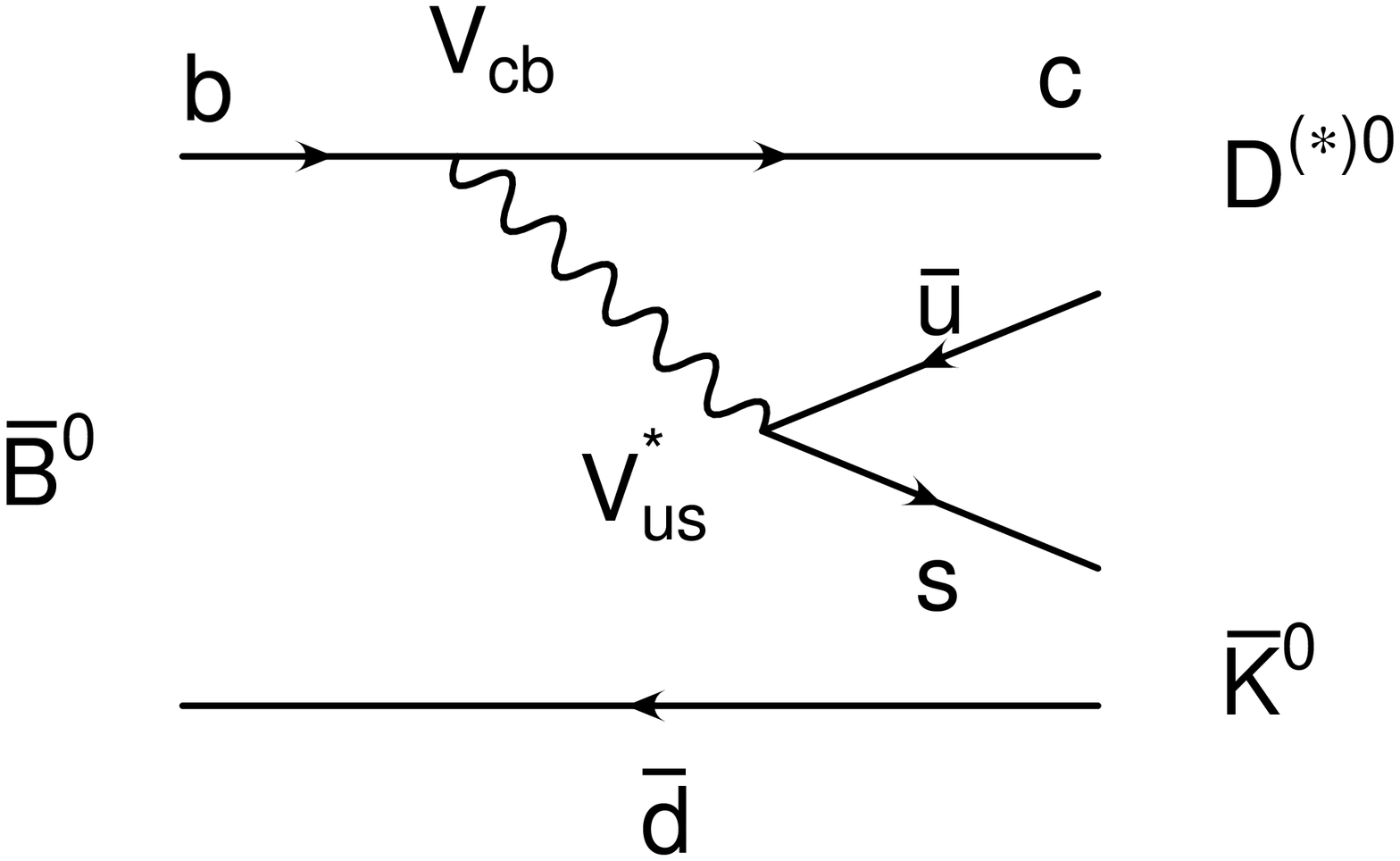,width=0.49\linewidth}
\epsfig{file=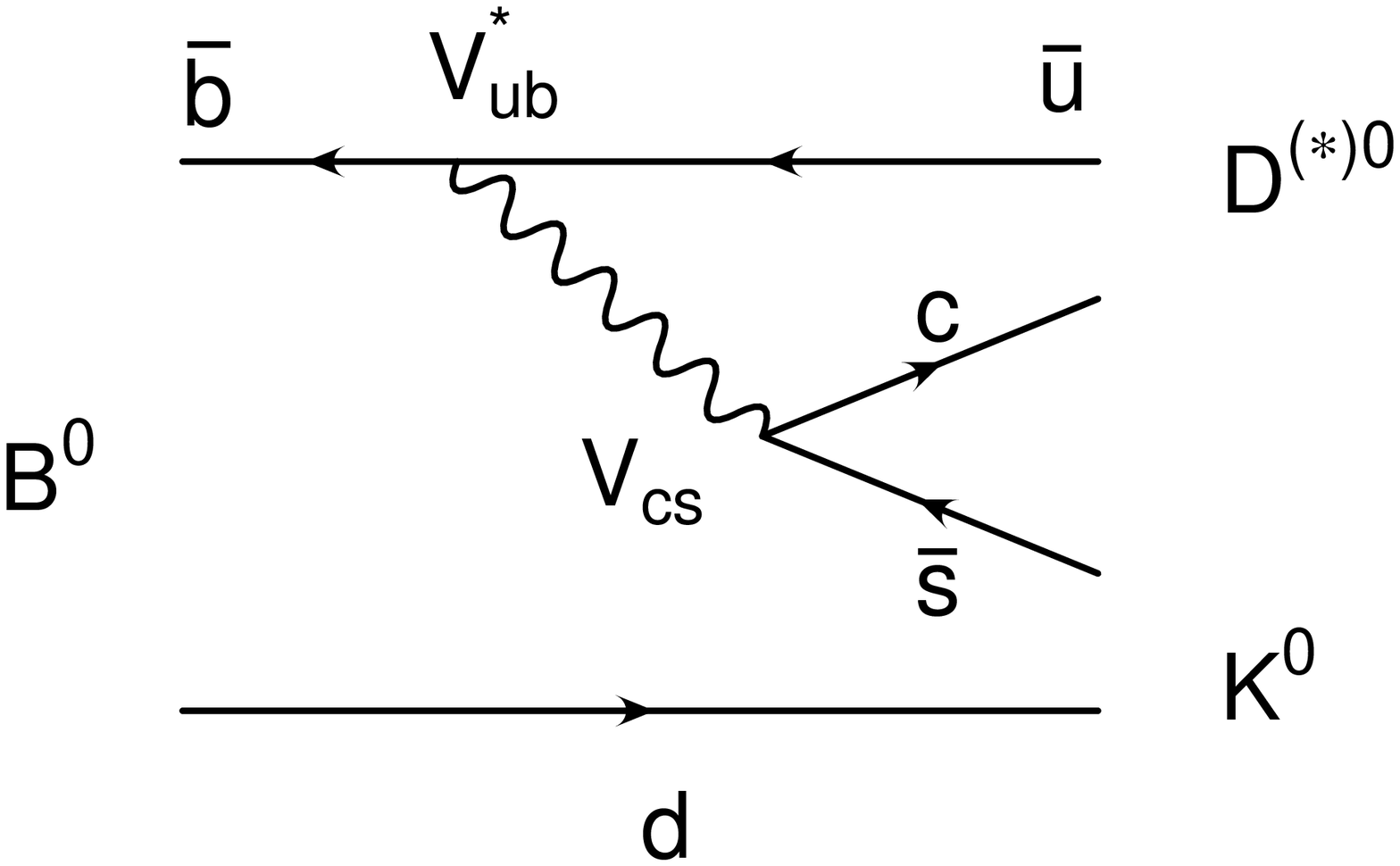,width=0.49\linewidth}
\end{center} 
\caption{
The decay diagrams for the $b\ra c$ transition $\Bzb\ra\DDstarz\Kzb$
and the $\bar b\ra \bar u$ transition $\Bz\ra\DDstarz\Kz$.
}
\label{fig:feyn}
\end{figure}

\begin{center}
\small

The \babar\ Collaboration,
\bigskip

%% author list as of 02-Jul-2004 (609 authors)
%
B.~Aubert,
R.~Barate,
D.~Boutigny,
F.~Couderc,
J.-M.~Gaillard,
A.~Hicheur,
Y.~Karyotakis,
J.~P.~Lees,
V.~Tisserand,
A.~Zghiche
\inst{Laboratoire de Physique des Particules, F-74941 Annecy-le-Vieux, France }
% ...

\end{center}\newpage
%%%%%%%%%%%%%%%%%%%%%%%%%%%%%%%%%%%%%%%%%%%%%%%%%%%%%%%%%%%%%%%%%%%%%%%%%%%%%%%
%  please do not use more than 80 columns which                              80
%%%%%%%%%%%%%%%%%%%%%%%%%%%%%%%%%%%%%%%%%%%%%%%%%%%%%%%%%%%%%%%%%%%%%%%%%%%%%%%
% theory introduction
With the discovery of \CP\ violation in the decays of neutral $B$ 
mesons~\cite{s2b-discovery} and
the precise measurement~\cite{sin2beta} of the angle $\beta$ of the 
Cabibbo-Kobayashi-Maskawa~(CKM) Unitarity
Triangle~\cite{CKM}, the experimental focus has shifted
%of the \CP-violation searches
toward the measurements of the angles $\alpha$ and $\gamma$.
Several methods have been suggested to measure $\gamma$ with small
uncertainties, but they all require large samples of $B$ mesons not
yet available.
The decay modes $\Bzb\ra\DDstarz\Kzb$ offer a new
approach for determination of \stwobg\ from measurement of
time-dependent \CP\ asymmetries in these decays~\cite{theory-DK}.
The \CP\ asymmetry appears as a  result
of interference between two possible diagrams leading
to the same final state $\DDstarz\KS$~(Figure~\ref{fig:feyn}):
a $b\ra c$ transition $\Bzb\ra\DDstarz\Kz$ and a $\bar b\ra \bar u$ transition
$\Bz\ra\DDstarz\Kz$.
A $\Bzb$ meson can either decay to a 
$\DDstarz\Kzb$~($\DDstarzb\Kzb$)  final state, or oscillate into a 
$\Bz$ which then decays to a $\DDstarz\Kz$~($\DDstarzb\Kz$)
final state.

The $\Bzb\Bz$ oscillation provides the weak phase $2\beta$ and 
the relative phase of the two decay diagrams is $\gamma$.

The sensitivity of this method depends on the ratio
$r\equiv|{\mathcal A}(\Bzb\ra\DDstarzb\Kzb)|/
|{\mathcal A}(\Bzb\ra\DDstarz\Kzb)|$~\cite{theory-DK}
of the decay amplitudes.
The ratio of the CKM matrix elements in the two amplitudes naively 
suggests $r\approx0.4$,
however, the $B$ decay dynamics can modify this expectation.
The ratio $r$ can be probed by measuring the rate for the decays
$\Bzb\ra\Dz\Kstarzb$ and 
$\Bzb\ra\Dzb\Kstarzb$, using the self-tagging decay $\Kstarzb\ra\Km\pip$.

The $\Bzb\ra\Dz\Kstarzb$ and  $\Bzb\ra\Dzb\Kstarzb$ decays are separated
by means of the correlation between the charge of the kaons
produced in the \Dz\ and \Kstarz\  decays: in the former decay
the two kaons must have the same charge, while in the latter they
are oppositely charged.
In the $\Bzb\ra\DDstarz\Kzb$ decays, the strangeness content of the
\Kzb\ is hidden and  one cannot distinguish between $\Bzb\ra\DDstarz\Kzb$ 
and $\Bzb\ra\DDstarzb\Kzb$. Hence in the remainder of this paper we refer 
to these decays as $B\ra\DDstarz\Kzb$.
We estimate the branching fractions ${\mathcal B}(B\ra\DDstarz\KKstarzb)$
from the measured color-suppressed decays 
$\Bzb\ra\DDstarz\piz$~\cite{color-suppressed} to be approximately
$3\times 10^{-5}$.
%
%In this Letter,  we present
In this paper,  we present
the first evidence for the decay $B \ra\Dstarz\Kzb$,
measurements of the  branching fractions 
${\mathcal B}(B \ra\Dz\Kzb)$ and ${\mathcal B}(\Bzb\ra\Dz\Kstarzb)$,
and a $90\%$ C.L. upper limit for the branching fraction of the 
$b\ra u$ transition $\Bzb\ra\Dzb\Kstarzb$.

%%% Data set
Results presented here are based on a sample of 
\BBcount\ million  $\Y4S \ra B \bar B$ decays collected with the \babar\
%\lumi~\invfb\ collected with the \babar\
detector between 1999 and 2003 at the \pep2\ asymmetric-energy
\epem\ collider operating near the \Y4S\ resonance.
The properties of the continuum $\epem\ra q\bar q \ (q=u,d,s,c)$ events
are studied with  a data sample of \offlumi~\invfb\ recorded 40~\mev\ 
below the \Y4S\ resonance. We also use large samples of simulated
$\Y4S\ra B\bar B$ and $\epem\ra q\bar q$ events which are about
three and 15 times the size of the data, respectively.

%%% Detector
%
% 1st version circulated to review committe
%
The \babar\ detector is described elsewhere~\cite{babar-detector-nim}. 
Only detector components relevant for this analysis are summarized here.
Trajectories of charged particles are measured in
a spectrometer consisting of a five-layer silicon vertex tracker 
(SVT) and a 40-layer drift chamber (DCH) operating in a 1.5~T axial 
magnetic field.  Charged particles
are identified as pions or kaons using information
from a detector of internally reflected Cherenkov
light, as well as measurements of energy loss from ionization
($dE/dx$) in the SVT and the DCH. 
Photons are detected using an electromagnetic calorimeter (EMC)
constructed of 6580 thallium-doped CsI crystals.

%%% Event Selection
%Event selection\\
We reconstruct the decays $\Bzb\ra\Dz\Kzb$, $\Dstarz\Kzb$, $\Dz\Kstarzb$,
 and $\Dzb\Kstarzb$ in the decay chains:
$\Dstarz\ra\Dz\piz$; $\Dz\ra\KPi$, \KPiPiz, and \KPiPiPi;
$\Kz\ra\KS\ra\pipi$; $\Kstarz\ra\Kp\pim$; and 
$\piz\ra\gamma\gamma$ 
(throughout this 
%
%Letter
paper 
charge-conjugated
decay modes are implied unless explicitly specified).
For each decay channel, the  optimal selection criteria are determined 
by maximizing the ratio  $S/\sqrt{S+B}$, where S and B are, respectively,
 the estimated signal and background  yields in simulated events.
%
%%% calibration Sample
A large sample of the more abundant $\Bp\ra\Dzb\pip$ decays,
in which the $\Dzb$ decays to the  $K^+\pi^-,\ K^+\pi^-\pi^0,$
and $K^+\pi^-\pi^+\pi^-$ final states, is used as a calibration sample
to measure efficiencies and resolutions for the selection variables.

Charged tracks used to  reconstruct \Dz\ and \Kstarz\ candidates
are required to have transverse momentum $p_T>100$~\mevc, and the 
\Kpm\ candidates must satisfy kaon identification criteria. 
These identification criteria have an average efficiency of about 90\%
while the probability of pions being mis-identified as kaons varies between
a few percent and 15\%.
The photons are reconstructed from clusters in the electromagnetic calorimeter 
with energy greater than 30~\mev\ that are consistent with photon showers.
We select \piz\ candidates from pairs of photon candidates and require
$115~\mevcc<m(\gamma\gamma)<150$~\mevcc.

The \KS\ candidates are selected from pairs of oppositely charged tracks
with invariant mass within 7~\mevcc~($2\sigma$) of the nominal \KS\ 
mass~\cite{PDG}.
The displacement $l_{\KS}$ of the \KS\ decay vertex from the interaction point
in the plane perpendicular to the beam axis divided by its measured
uncertainty $\sigma_{\KS}$ must be greater than 2.
The \Kstarz\ candidates are selected from pairs of oppositely charged
\Kp\ and \pim\ tracks, with invariant mass $m(\Kp\pim)$ within 50~\mevcc\ 
of the nominal \Kstarz\ mass~\cite{PDG}.
We also require  $|\cos{\theta_{h}}|>0.4$, where
$\theta_h$ is the \Kstarz\ helicity angle, 
defined as the angle between the direction of the \Kstarz\ in the \Bz-meson
rest frame and the direction of its daughter \Kp\ in the \Kstarz\ rest frame.
For signal $\Bzb\ra\Dz\Kstarzb$ candidates, $\theta_{h}$ follows a
$\cos^2\theta_{h}$ distribution
while the combinatorial background is distributed uniformly.

We reconstruct the \Dz\ candidates in the \KPi\ and \KPiPiPi\ 
decay modes by combining charged tracks.
Combinations with an invariant mass within $2\sigma$
of the nominal \Dz\ mass $m_{\Dz}$ are retained.
In the $\Dz\ra\KPiPiz$ selection, the \piz\ candidates are required to have
a center-of-mass (CM) momentum  $p_{\piz}^*$ greater than $400$~\mevc.
For each $\KPiPiz$ combination,
we use the kinematics of the decay products and the known 
properties of the Dalitz plot for
%resonant sub-structure of
this decay~\cite{dalitz} to compute the square of the decay amplitude $|A|^2$.
We select combinations with
$|A|^2$ greater than $5\%$ of its maximum value. 
This requirement selects mostly the $K \rho$ region of the Dalitz plot.
It rejects $62\%$ of the combinatorial background
and has an efficiency of $76\%$, as measured with the $\Bp\ra\Dzb\pip$
calibration sample.
The combinations with the invariant mass within 25~\mevcc~($2.5\sigma$)
of $m_{\Dz}$ are retained.
For the purpose of cross checks, we define a \Dz\ mass sideband
as the region between 45 and 90~\mevcc\ away from $m_{\Dz}$
 in the $\Dz\ra\KPi$  and \KPiPiPi\ mode, and between 85 and 
160~\mevcc\ away from $m_{\Dz}$ in $\Dz\ra\KPiPiz$.

The \Dstarz\ candidates are selected from combinations of a \Dz\ and 
a \piz\ with $p_{\piz}^*>70$~\mevc.
After kinematically constraining \Dz\ and \piz\ candidates to their
nominal masses, we select the candidates with the mass difference 
$\Delta m \equiv |m(\Dstarz)-m(\Dz)-142.2~\mevcc| <3.3$~\mevcc.

Two standard kinematic variables are used to select \Bz\ candidates:
the energy-substituted mass 
$\mes\equiv\sqrt{(\frac{1}{2}s+\mathbf{p}_0\cdot\mathbf{p}_{B})^2/E_0^2-\mathbf{p}_B^2}$ 
and the energy difference $\DeltaE\equiv E_B^*-\frac{1}{2}\sqrt{s}$, where
the asterisk denotes the CM frame, 
$s$ is the square of the total energy in the CM frame,
$\mathbf{p}$ and $E$ are, respectively, three-momentum and energy, and
the subscripts $0$ and $B$ refer to \Y4S\ and \Bz, respectively.
For signal events, $\mes$ is centered around the $\Bz$ mass 
with an r.m.s. resolution of about 2.6~\mevcc,
dominated by the knowledge of the $e^+$ and $e^-$  beam energies.
The \DeltaE\ resolution is dominated
by the momentum and energy resolutions of the detector and hence varies
for different decay modes.
We constrain the mass of the \DDstarz\ and \KS\ candidates to their
respective nominal values. In simulated events the r.m.s. \DeltaE\
resolution is found to be $\approx13$~\mev\ for all \Bz\ decay modes. 
The \Bz\ candidates are required to have \DeltaE\ within 30~\mev\ of
the mean value measured in the $\Bp\ra\Dzb\pip$ calibration sample.

%%% Continuum Suppression
We use two variables to reject most of the remaining
background, which is dominated by continuum events:
the polar angle $\theta^*_B$ of the \Bz\ candidate in the CM frame and
a Fisher discriminant~\cite{fisher} based on the energy flow in the
rest of the event, after removing the \Bz\ decay products.
The Fisher discriminant is defined as a linear combination of 
$|\cos{\theta_{TB}^*}|$, where $\theta_{TB}^*$ is the angle in the CM frame
between the thrust axis of the \Bz\ and that of the remaining charged and
neutral particles in the rest of the event, and the Legendre monomials
${\mathcal L}^0$ and ${\mathcal L}^2$ defined as 
${\mathcal L}^i\equiv\sum_{j} p_{j}^*\ \cos^{i}{\theta_j}$; here
$p^*_j$ is the CM momentum and $\theta_j$ the angle between the
direction of remaining particles in the event¿½
with respect to the thrust axis of the \Bz\ candidate.
The requirement on \fisher\ varies for each decay channel
because of different levels of the background. 
In the $\DDstarz\KS$ and $\Dz\Kstarzb$ final states our
requirement has a typical signal efficiency of about
80\% while rejecting about 85\% of the background.
A tighter requirement in the $\Bzb\ra\Dzb\Kstarzb$ mode
rejects 95\% of the background and has a signal efficency
of 55\%.
The requirements on angle $\theta^*_B$ are 
$|\cos\theta^*_B|<0.75$ for $\Bzb\ra\Dzb\Kstarzb$, and
$|\cos\theta^*_B|<0.85$ for all other decay modes.

%%% Multiple candidates
In about 2--3\% of the events more than one \Bz\ candidate
with $\mes>5.2$~\gevcc\  satisfy the selection criteria.
In the $\Dz\Kzb$, $\Dz\Kstarzb$, and $\Dzb\Kstarzb$ final states
the candidate with the smallest \DeltaE\ is selected. In the
$\Dstarz\KS$ final state, we pick the candidate with the smallest $\chi^2$ 
computed from the measured value of $m(\Dz)$ and $m(\Dstarz)-m(\Dz)$,
their nominal values, and their resolutions in data.

%%% m_ES fit
The signal yield for each \Bz\ decay mode is determined with a
binned maximum likelihood fit to the \mes\ distribution
for each \Dz\ decay mode. The distribution is modeled with
a Gaussian for the signal and a threshold function for the combinatorial
background.
The mean and the r.m.s. resolution of the Gaussian are fixed to values
measured in the  $\Bp\ra\Dzb\pip$ calibration sample.
The threshold function parameterizing the background is defined as
${\mathcal A}(\mes)\sim \mes \sqrt{1-x^2}\exp\{-\xi(1-x^2)\}$~\cite{argus}, 
where $x=2\mes/\sqrt{s}$ and $\xi$ is a shape parameter.

\begin{figure}[htb]
\begin{center}
\epsfig{file=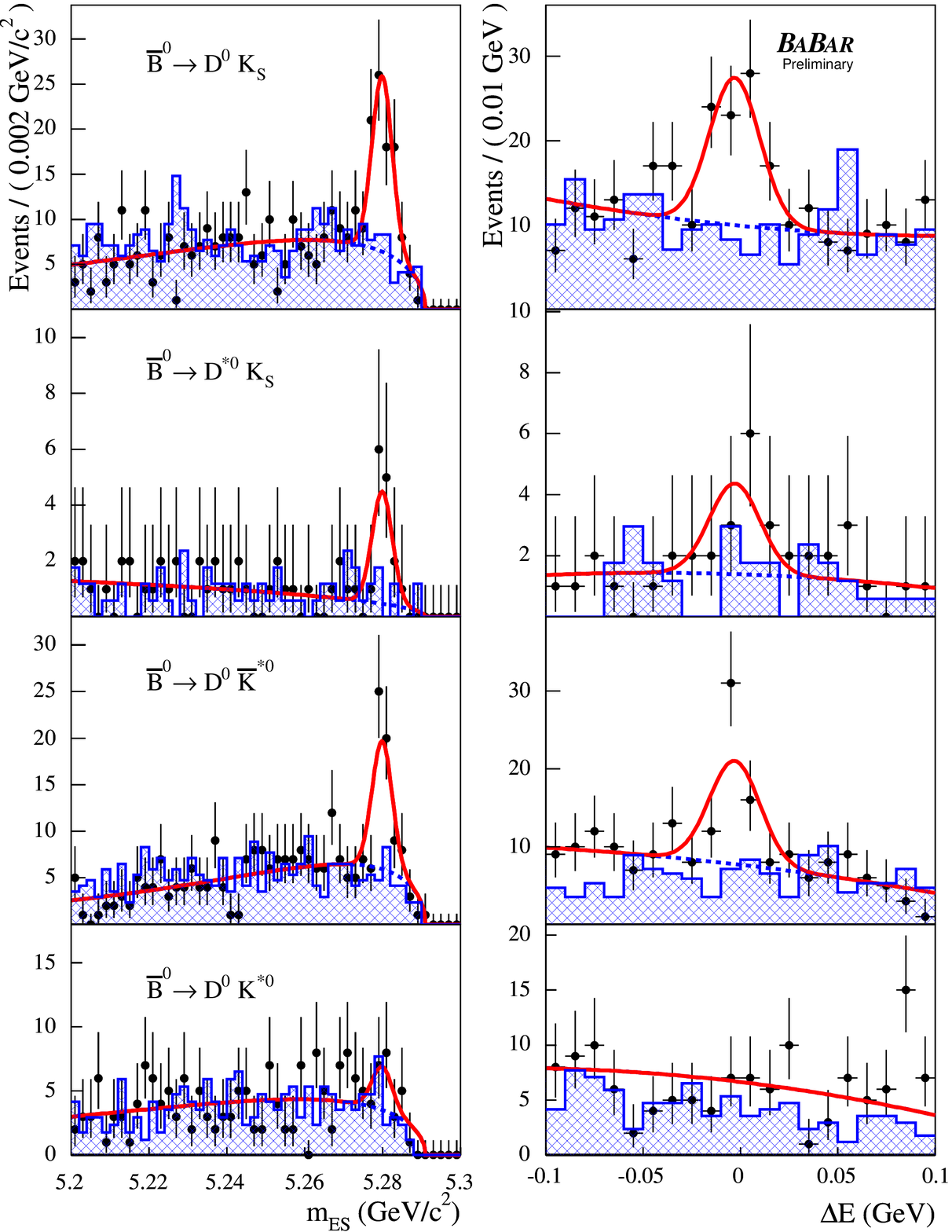,width=0.95\linewidth}
\end{center} 
\caption{
Distribution of \mes\ (left) and \DeltaE\ (right)
for $B\ra\DDstarz\Kzb$ and $\Bzb\ra\Dz\Kstarzb$ and $\Dzb\Kstarzb$ candidates.
The points are the data, the curve is the result of the fit, and the 
hatched histogram is the distribution of candidates in \Dz\ sidebands.
}
\label{fig:yields}
\end{figure}

\begin{figure}[htb]
\begin{center}
\epsfig{file=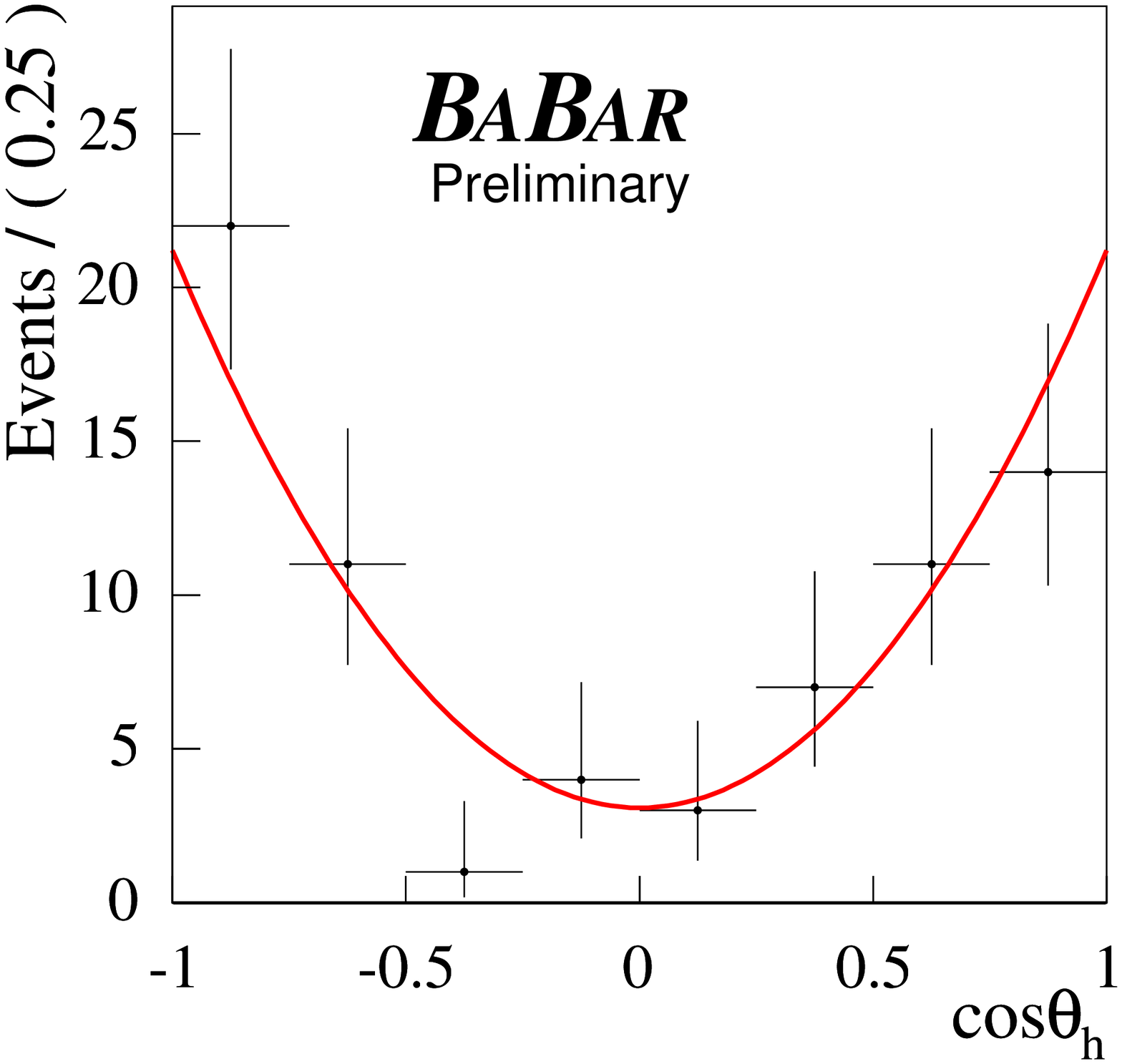,width=0.9\linewidth}
\end{center} 
\caption{
Distribution of $\cos\theta_h$ for selected $\Bzb\ra\Dz\Kstarzb$ candidates.
The points are data, after subtracting the combinatorial background, and the
curve is the result of a fit, which is consistent with the expected
$\cos^2\theta_h$ distribution.
}
\label{fig:hel K*0}
\end{figure}

%%% Yields table and figure
The measured signal yields are summarized in Table~\ref{tab:yields},
and the \mes\ distributions for the sums of all three \Dz\ decay modes
are illustrated in Figure~\ref{fig:yields}.
%%%%%%%
\begin{table}[!t]
\begin{center}
%                 1234567
\begin{tabular}{lccccc}
\hline
  \Dz\ channel & $N_{S}$  & & $N^{\rm pk}$ &
   $\varepsilon_{\rm eff}~(10^{-3})$ & ${\mathcal B}~(10^{-5})$  \\
\hline
       & \multicolumn{5}{c}{$B\ra\Dz\Kzb$} \\
\hline
\KPi    &   $18.9\pm5.6$ & & $2.1\pm1.1$ & $2.85$ &  $4.8\pm1.6\pm0.5$ \\
\KPiPiz &   $18.0\pm6.3$ & & $0.9\pm0.5$ & $2.20$ &  $6.3\pm2.3\pm0.7$ \\
\KPiPiPi&   $26.8\pm6.9$ & & $0.0\pm1.1$ & $2.10$ & $10.3\pm2.7\pm0.9$ \\
 All    &   $64\pm11   $ & & $2.1\pm1.1$ & $7.15$ &  $6.2\pm1.2\pm0.4$ \\
\hline
       & \multicolumn{5}{c}{$B\ra\Dstarz\Kzb$} \\
\hline
\KPi    &   $3.1\pm2.0$ & & $0.3\pm0.2$ & $0.80$ & $3.3\pm2.2\pm0.6$ \\
\KPiPiz &   $4.9\pm2.7$ & & $0.2\pm0.2$ & $0.51$ & $8.5\pm4.9\pm1.4$ \\
\KPiPiPi&   $3.2\pm2.0$ & & $0.1\pm0.1$ & $0.34$ & $8.6\pm5.5\pm1.0$ \\
 All    &  $11.2\pm3.9$ & & $0.8\pm0.4$ & $1.66$ & $4.5\pm1.9\pm0.5$ \\
\hline
       & \multicolumn{5}{c}{$\Bzb\ra\Dz\Kstarzb$}  \\
\hline
\KPi    &  $17.7\pm5.2$ & & $1.6\pm0.8$ & $2.34$ & $6.6\pm2.1\pm0.8$ \\
\KPiPiz &  $11.7\pm5.5$ & & $1.0\pm0.5$ & $2.09$ & $5.2\pm2.6\pm0.7$ \\
\KPiPiPi&  $14.9\pm5.4$ & & $1.5\pm0.8$ & $2.10$ & $6.7\pm2.7\pm1.0$ \\
 All    &  $45.2\pm9.2$ & & $4.4\pm2.2$ & $6.54$ & $6.2\pm1.4\pm0.6$ \\
\hline
       & \multicolumn{5}{c}{$\Bzb\ra\Dzb\Kstarzb$}  \\
\hline
 All    &  $11.0\pm5.9$ & & $5.5\pm4.5$ & $4.43$ & $<\dzkstarzLim$ 90\% C.L. \\
\hline
\end{tabular}
\caption{
Signal yield $N_S$, estimated peaking background $N^{\rm pk}$,
effective signal efficiency $\varepsilon_{\rm eff}$, and the measured branching
fraction ${\mathcal B}$ for the $B\ra\DDstarz\KKstarzb$ decays.
The efficiency $\varepsilon_{\rm eff}$ is defined as $\varepsilon\times{\rm BF}$, where
$\varepsilon$ is the signal reconstruction efficiency and ${\rm BF}$ are the appropriate
intermediate  branching fractions  for \Dstarz, \Dz, \Kstarz, and \Kz\ decays
to final states reconstructed in this analysis.
}
\label{tab:yields}
\end{center}
\end{table}
%%%%%
%
%
The signal yields measured from the \DeltaE\ distribution, 
after removing the requirement on \DeltaE\ and selecting
candidates with $5.273~\gevcc <\mes <5.288$~\gevcc, 
are found to be in good agreement with the \mes\ yields.
In this case, for the small fraction of events with more than one selected
candidates, we choose the one with \mes\ closest to the $B$ mass.
The \DeltaE\ distribution for the signal is modeled with a Gaussian and a 
second-order polynomial is used for the background.
The \DeltaE\ distributions for the selected events are 
also shown in Figure~\ref{fig:yields}.
The combinatorial background in both \mes\ and \DeltaE\ distributions
are described well by events in the sidebands of the \Dz\ mass, 
which are shown as hatched histograms in Figure~\ref{fig:yields}.
As a further cross check we 
examine the \Kstarz\ helicity angle $\theta_{h}$ of the $\Bzb\ra\Dz\Kstarzb$ 
candidates with $5.273~\gevcc <\mes <5.288$~\gevcc\
after removing the requirement $|\cos\theta_{h}|>0.4$.
Figure~\ref{fig:hel K*0} shows
the distribution of $\cos{\theta_{h}}$ for these candidates, after
subtracting the combinatorial background from the \Dz\ sidebands.
We fit this spectrum with a sum of a flat component for the background
and a $\cos^{2}\theta_{h}$ distribution for the signal, and find the
fraction of the former to be consistent with zero as expected.

%%% Peaking backgrounds
The measured signal yields $N_S$ include a small  contribution from other
$B$ decays that can be mis-reconstructed as signal events and 
accumulate near the $B$ mass.
We have studied the contribution of such potential ``peaking''
backgrounds with large samples of simulated events, corresponding to 
typically between
100 and 1000 times the size of our data,  for the following categories
of decays:
$(1)$ $\Bz\ra\Dzb\rho^0,\rho^0\ra\pip\pim$ decays, where one of the
two pions is mis-identified as a charged kaon; 
$(2)$ the $\Bz\ra\Dm\pip$ decays, followed by the Cabibbo-suppressed decays
$\Dm\ra\KKstarz\Km$ and
$\Bz\ra\Dm\Kp$, $\Dm\ra\KKstarz\pim$ reconstructed in the
$\Dz(\Km\pip)\KKstarz$ final states; 
$(3)$ charmless $\Bz\ra\Kp\pim\KS(n\pi)$ where the \Kp\ and \pim\ are
wrongly combined to form a $\Dzb\ra\Kp\pim$ candidate;
$(4)$ $\Bz\ra\Dstarz\KKstarzb$,
$\Dstarz\ra\Dz\gamma$ candidates, where a low-energy photon $\gamma$ is
not reconstructed;
$(5)$ the decays $\Bp\ra\Dstarzb\Kp, \Dstarzb\ra\Dzb\piz/\gamma$, 
$\Bp\ra\Dzb\Kstarp,\Kstarp\ra\Kp\piz,\KS\pip$, and 
$\Bz\ra\Dstarm\Kp, \Dstarm\ra\Dzb\pim$, where a low-energy \piz\ or
photon is replaced by a random low-momentum charged track.
The contribution of category $(1)$ is found to be less than 0.01
events and is hence neglected.  The contribution of category $(2)$
is also negligible in all modes, except for $B\ra\Dz\Kzb, \Dz\ra\Km\pip$.
We eliminate 87\% of these events by requiring the invariant masses 
$m(\KS\Km)$ and $m(\KS\pip)$ to be more than 20~\mevcc\ away from 
the nominal \Dp\ mass.
The \mes\ spectrum of the remaining background events in this category
and in  categories $(3)$--$(5)$
shows a broad enhancement near the $B$ mass.
The contributions of these events to the signal yields are measured
by performing a Monte Carlo study and  are summarized
in Table~\ref{tab:yields}. We assign a 50\% systematic error to the estimated
peaking background contribution due to the uncertainty on the branching
fractions of some of these $B$ decays.
In the decay $\Bzb\ra\Dzb\Kstarzb$, the charge correlation
used in  the selection removes all contributions from known $B$ decays
included in  our simulation.  We 
estimate the peaking background for this decay mode from the
\Dz\ sidebands to be $5.5\pm4.5$ events.

%%% Signal significance
The significance of the signal yields is determined
by taking into account the sum $N_B$ of the combinatorial and peaking 
backgrounds and its uncertainty $\delta N_B$. We generate
one million experiments where the expected number of background $N_B^i$
for the $i$-th experiment is extracted from a Gaussian with mean $N_B$ 
and width  $\delta N_B$.
Assuming a Poisson distribution for the background events, the significance
is determined from the fraction of experiments with $N_B^i>N_{\rm obs}$
with $N_{\rm obs}$ being the number of candidates in data with 
$5.273~\gevcc<\mes < 5.288$~\gevcc. The $B\ra\Dstarz\Kzb$ signal has a 
significance
of $3.3\sigma$ and is the first evidence for this decay mode. The
significance of the $\Bzb\ra\Dz\Kstarzb$ and $\Bzb\ra\Dzb\Kstarzb$
are, respectively, $4.8\sigma$ and $2.0\sigma$.
For the $\Dz\Kzb$ final state, where the number of observed events
is large, we estimate  the signal significance to be $5.8\sigma$ from the 
measured signal yield and its uncertainty.

%%% Systematic errors
The systematic uncertainties for the branching fractions are reported
in Table~\ref{tab:yields} and include contributions  from 
estimated peaking background~(3--11\%),
fit parameters~(2--8\%), 
\DDstarz\ branching fraction~(2.4--6.9\%), 
\piz\ reconstruction efficiency~(2.5\% per photon),
charged-track reconstruction efficiency~(0.8\% per track), 
Monte Carlo statistics~(1--4\%),
efficiency correction factors~(1--4\%),
kaon identification~(2\% per kaon), 
\KS\ reconstruction efficiency~(1.6\%),
and number of $B\bar B$ events~(1.1\%).

%%% Branching fraction
The branching fraction ${\mathcal B}$ of each \Bz\ decay mode is computed
as the weighted average of the branching fractions  $B_{j}$ 
in each \Dz\ channel $D_j=\{\KPi,\KPiPiPi,\KPiPiz\}$, computed as
\begin{eqnarray}
{\mathcal B}_{j}  = \frac{ N_{S_{j}} - N_{j}^{\rm pk}  }
                       { N_{B\bar B} \times 
                         {\mathcal B}_{D_j} \times
                         {\mathcal B}_K \times
                         \varepsilon_{j} \nonumber
                       }
\label{eq:br expr}
\end{eqnarray}
where $N_{S_{j}}$ is the signal yield from the \mes\ fit, 
$N_{j}^{\rm pk}$ is the estimated peaking background from 
Table~\ref{tab:yields}, 
$N_{\B\bar B}$ is the total number of $\Y4S\ra B \bar B$ events, 
${\mathcal B}_{D_j}$ is the branching fraction ${\mathcal B}(\Dz\ra D_j)$
in  $B\ra\Dz\KKstarzb$ and 
${\mathcal B}(\Dstarz\ra\Dz\piz)\times{\mathcal B}(\Dz\ra D_j)$ in
$B\ra\Dstarz\Kzb$, 
${\mathcal B}_K $ is the known $\Kz\ra\KS\ra\pip\pim(\Kstarz\ra\Kp\pim)$
branching fraction
in
$B\ra\DDstarz\Kzb(\Kstarzb)$,
and $\varepsilon_{j}$ is the signal reconstruction efficiency.
We assume ${\mathcal B}(\Y4S\ra\Bz\Bzb)=0.5$.
%

%%% Results
We measure
\begin{eqnarray}
\BR(B\ra\Dz\Kzb) &=& (\dzkzBrVal\pm\dzkzBrStat
\pm\dzkzBrSyst)\brscale \nonumber \\
\BR(B\ra\Dstarz\Kzb) &=& (\dstarzkzBrVal \pm \dstarzkzBrStat
\pm \dstarzkzBrSyst)\brscale \nonumber  \\
\BR(\Bzb\ra\Dz\Kstarzb) &=& (\dzbkstarzBrVal \pm \dzbkstarzBrStat
\pm\dzbkstarzBrSyst)\brscale \nonumber \\
\BR(\Bzb\ra\Dzb\Kstarzb) &=& (\dzkstarzBrVal \pm \dzkstarzBrStat
\pm \dzkstarzBrSyst)\brscale \nonumber
\end{eqnarray}
where the uncertainties are, respectively, statistical and systematic.
For the decay  $\Bzb\ra\Dzb\Kstarzb$ we use the Bayesian method to compute 
the upper limit $N_{UL}$ on the observed number of events at 90\% confidence 
level as $\int^{N_{UL}}_0  {\mathcal L}(N)\ dN=0.9$, where
${\mathcal L}(N)$ is the binned maximum  likelihood function
from the fit to the \mes\ distribution. We assume a flat prior
probability density function for ${\mathcal B}>0$.
After accounting for the systematic uncertainties we obtain at 90\% C.L.
$\BR(\Bzb\ra\Dzb\Kstarzb) < \dzkstarzLim\times 10^{-5}$.

%%% Summary
In summary, we have presented  evidence 
for the decay $\Bzb\ra\Dstarz\Kzb$ as well as new measurements
of the branching fractions for the decays  $\Bzb\ra\Dz\Kzb$ and $\Dz\Kstarzb$.
Our measurements are in agreement with the expectation 
derived from Ref.~\cite{color-suppressed} and with previous 
measurements~\cite{belle-d0ks-prl}.
We use the central value of our measurement for
$\BR(\Bzb\ra\Dzb\Kstarzb)$ and obtain
$r < 0.8$ at the 90\% C.L. from a central value of
$r=0.4\pm0.2\ {\rm (stat.)}\pm0.2\ {\rm (syst.)}$. 
The main contribution to the systematic uncertainty is from the estimated 
peaking background since most systematic uncertainties on the branching 
fractions cancel in the ratio.

We are grateful for the excellent luminosity and machine conditions
provided by our \pep2\ colleagues, 
and for the substantial dedicated effort from
the computing organizations that support \babar.
The collaborating institutions wish to thank 
SLAC for its support and kind hospitality. 
This work is supported by
DOE
and NSF (USA),
NSERC (Canada),
IHEP (China),
CEA and
CNRS-IN2P3
(France),
BMBF and DFG
(Germany),
INFN (Italy),
FOM (The Netherlands),
NFR (Norway),
MIST (Russia), and
PPARC (United Kingdom). 
Individuals have received support from CONACyT (Mexico), A.~P.~Sloan Foundation, 
Research Corporation,
and Alexander von Humboldt Foundation.

\end{document}